\documentstyle[PASJadd,epsf]{PASJ95}
\newcommand{\vol}{00}
\newcommand{\no}{0}
\newcommand{\lett}{}
\newcommand{\spage}{L1}
\newcommand{\rdate}{}
\newcommand{\adate}{}
\begin{document}

\title{
Substructures Revealed by the Sunyaev--Zel'dovich Effect at 150~GHz
in the High Resolution Map of RX~J1347--1145
}

\author{
Eiichiro {\sc Komatsu},$^1$ $^2$
Hiroshi {\sc Matsuo},$^3$
Tetsu {\sc Kitayama},$^4$
Makoto {\sc Hattori},$^2$
Ryohei {\sc Kawabe},$^5$\\
Kotaro {\sc Kohno},$^5$
Nario {\sc Kuno},$^5$
Sabine {\sc Schindler},$^6$
Yasushi {\sc Suto},$^7$ $^8$
and Kohji {\sc Yoshikawa}$^9$
\\[12pt]
$^1${\it Department of Astrophysical Sciences,
Princeton University, Princeton, NJ 08544, USA}\\
{\it E-mail(EK): komatsu@astro.princeton.edu}\\
$^2${\it Astronomical Institute, T\^{o}hoku University, Aoba,
Sendai 980-8578}\\
$^3${\it National Astronomical Observatory, Mitaka, Tokyo 181-8588}\\
$^4${\it Department of Physics, Tokyo Metropolitan University,
Hachioji, Tokyo 192-0397}\\
$^5${\it Nobeyama Radio Observatory, Minamimaki,
Minamisaku, Nagano 384-1305}\\
$^6${\it Astrophysical Research Institute, Liverpool John Moores
University,
Byrom Street, Liverpool, L3 3AF, England, UK}\\
$^7${\it Department of Physics, The University of Tokyo,
Tokyo 113-0033}\\
$^8${\it Research Center for the Early Universe,
School of Science,
The University of Tokyo, Tokyo 113-0033}\\
$^9${\it Department of Astronomy, Kyoto University, Kyoto
606-8502}
}

\abst{ We report on mapping observations toward the region of the most
  luminous X-ray cluster RX~J1347--1145 ($z=0.45$) through the
  Sunyaev--Zel'dovich effect at 21~GHz and 150~GHz with the Nobeyama
  45-m telescope. While a low angular resolution image at 21~GHz
  (beam-size $\sigma_{\rm FWHM}$ of $76''$) shows a consistent feature
  with the ROSAT/HRI X-ray image, a higher angular resolution image
  ($\sigma_{\rm FWHM} =13''$) at 150~GHz reveals complex morphological
  structures of the cluster region, which cannot be simply described by
  the spherical isothermal $\beta$-model.  If such inhomogeneous
  morphological features prove to be generic for high redshift
  clusters, distance measurements to the clusters based on their
  Sunyaev--Zel'dovich data with low angular resolution imaging
  should be interpreted with caution.  }
\kword{cosmology: observations -- distance scale -- cosmic
  microwave background -- galaxies: clusters: individual
  (RX~J1347--1145) -- X-rays: galaxies}
\maketitle
\thispagestyle{headings}

\section{Introduction}

The multi-band observation of galaxy clusters up to $z \sim 1$
provides a unique opportunity to reconstruct the clustering
evolution, the cosmological parameters, and the peculiar velocity
field on large scales (e.g., Bahcall 1988; Rephaeli 1995; Birkinshaw
1999).  In particular, the recent mapping observations of clusters
using the state-of-the-art interferometers (Carlstrom, Joy,
Grego 1996; Carlstrom et al. 2000) are accumulating impressive
{\it negative} intensity images in centimeter bands
through the Sunyaev--Zel'dovich (SZ) effect (Zel'dovich, Sunyaev 1969).
The most important cosmological application of such data is to
estimate the global value of the Hubble constant, $H_0$. Although the
SZ effect can be used as a potential standard candle, previous
attempts to estimate $H_0$ were not sufficiently accurate (e.g.,
Kobayashi, Sasaki, Suto 1996; Birkinshaw 1999).  This is probably
because they usually neglect the non-sphericity, substructure, and
non-isothermal profile of the intra-cluster matter.  
Recent numerical simulations of clusters (e.g., Inagaki, Suginohara, Suto
1995; Yoshikawa, Itoh, Suto 1998; Jing, Suto 2000) have shown that the
departure from the spherical isothermal $\beta$-model is appreciable
and that the inhomogeneous morphology is quite generic.

High angular resolution imaging through the SZ effect is
essential in resolving detailed structure of clusters. In centimeter
interferometric measurements, the beam-size of $\sigma_{\rm FWHM}\sim
100''$ is typical, and can be as small as $\sim 40''$ (Carlstrom, Joy,
Grego 1996), while the brightness sensitivity is degraded.  In millimeter
and submillimeter bands, on the other hand, the sensitive higher
resolution imaging is now feasible with bolometer array detectors.
This observing technique has been successfully applied to the
brightest X-ray cluster RX~J1347--1145 ($z=0.451\pm 0.003$) at 350~GHz
with $\sigma_{\rm FWHM} =15''$ (Komatsu et al. 1999), and subsequently
at 143~GHz with $\sigma_{\rm FWHM} =23''$ (Pointecouteau et al. 1999).

The bolometric X-ray luminosity of this cluster is exceptionally high,
$L_{\rm X}h_{50}^{2}=2\times 10^{46}~{\rm erg~s^{-1}}$ (Schindler et
al. 1997), where $h_{50}$ is the Hubble constant in units of $50~{\rm
km~s^{-1}~Mpc^{-1}}$. 
Throughout this {\it Letter}, we use 
$\Omega_0=1$ and $\lambda_0=0$ for simplicity.  
The ASCA observation implies that the
emission-weighted temperature is $k_{\rm B}T_{\rm e}=9.3^{+1.1}_{-1.0}~{\rm
keV}$ (Schindler et al. 1997). 
Applying the spherical isothermal $\beta$-model
to the latest ROSAT/HRI data, we obtain the central
electron number density, 
$n_{\rm e0}h_{50}^{-1/2}=0.093\pm 0.004~{\rm cm^{-3}}$,
the core radius, $\theta_{\rm c}=8.\hspace{-2.5pt}''4\pm
1.\hspace{-2.5pt}''0$, and $\beta=0.57\pm 0.02$, where quoted errors
are 90\% confidence levels.  The total mass inferred from the hydrostatic
equilibrium is 
$M(<2~h^{-1}_{50}~{\rm Mpc})=1\times 10^{15}~h^{-1}_{50}~M_\odot$. 
This cluster does not follow the temperature--luminosity relation
because of the extremely high luminosity (Schindler 1999),
and is known to be the strong cooling-flow cluster (Schindler et al. 1997;
Allen, Fabian 1998).  In this sense, this cluster may not be typical,
but is an interesting target to be studied in detail individually.

The above two attempts to map the cluster using the SZ effect are not
sensitive enough to resolve detailed structures; the former data
(Komatsu et al. 1999) are very noisy because of the bad weather
condition, and the latter (Pointecouteau et al. 1999) mapped only the
central narrow stripe of the cluster. Therefore, as a part of the
multi-band observing project of RX~J1347--1145, we carried out the SZ
mapping observation of the cluster at 21~GHz and 150~GHz with the Nobeyama
45-m telescope. In this {\it Letter}, we report on
the complex morphological structures detected in the 150~GHz map with the
unprecedented angular resolution ($\sigma_{\rm FWHM}=13''$),
which can hardly be identified from the lower resolution map at 21~GHz
($\sigma_{\rm FWHM} =76''$). This demonstrates the importance of the
high angular resolution imaging of clusters through the SZ effect.

\section{The Sunyaev--Zel'dovich Mapping Observations
 with the Nobeyama 45-m telescope}

\subsection{Centimeter Mapping at 21~GHz}

Mapping observations at 21~GHz were carried out during February 16--27
and April 14--22, 2000, in a raster-scan mode using the
dual-polarization HEMT amplifier mounted on the Nobeyama 45-m
telescope.  In total, 9 scans were performed along the two orthogonal
directions and each scan was separated by $40''$ yielding the final
field-of-view of $6'\times 6'$.  A reference beam position was set to
be $400''$ away from a main beam, and the beam-size $\sigma_{\rm
  FWHM}=76.\hspace{-2.5pt}''5$ was estimated by observing 3C279. The
beam is accurately fitted to a Gaussian.  
The exposure time amounts to
31.2~ksec in February and 32.1~ksec in April.  System noise
temperatures were typically 135~K in February and 185~K in April.  We
calibrated the primary flux using 3C286 ($2.56\pm 0.02~{\rm Jy}$; Ott
et al. 1994). The stability of antenna efficiency was tracked by
monitoring the pointing source (1334-127), and we
found that an rms variation in peak flux was 2\% in February 
and 3\% in April.
A pointing offset is negligible compared 
with the beam-size.

We subtracted the low frequency scanning-noise from the map 
on the basis of the PLAIT method (Emerson, Gr\"ave
1988) with a scale-length half the scan-length.  The resulting
$1\sigma$ noise-levels in images are $1.3~{\rm mJy~beam^{-1}}$ in
February and $1.7~{\rm mJy~beam^{-1}}$ in April.  Combining the
February and April runs, the final image achieves the noise-level of
$0.9~{\rm mJy~beam^{-1}}$.  An absolute calibration error is 1\%,
which is dominated by the flux error of the primary calibrator 3C286.

\subsection{Millimeter Mapping at 150~GHz}

The higher angular resolution mapping of the cluster was performed
with the Nobeyama Bolometer Array (NOBA; Kuno et al. 1993) on March 16
(20~ksec) and April 15 (8.5~ksec), 1999, as well as during February
16--27, 2000 (52.7~ksec).  The total integration time is 81.2~ksec.
NOBA consists of seven bolometers in hexagonal pattern with their
band-passes centered at 150~GHz and bandwidths of 30~GHz. 
Bolometers are read-out through six differential circuits between 
a central bolometer and the other six surrounding ones. 
Fluctuations in atmospheric emissions are subtracted in real time 
by the readout electronics. The beam switching, or the sky chopping, 
is not used. The observation of this cluster was made with 
raster scans.  A position angle of the array to the scan direction is 
$19.\hspace{-2.5pt}^\circ1$. A single raster scan yields seven scan 
paths separated by $5.\hspace{-2.5pt}''3$ each, and an observing stripe
of $37.\hspace{-2.5pt}''1$ in width. 
Three stripes thus cover the field-of-view of 
$1.\hspace{-2.5pt}'9\times 1.\hspace{-2.5pt}'9$ in right ascension and 
declination.  Image restoration is performed using the six differential 
signals.

At the beginning and the end of each observing day, an elevation scan
was made to measure the atmospheric transparency. The measured zenith
optical depths at 150~GHz were 0.094--0.15 and 0.065--0.098 in our
1999 and 2000 runs, respectively.  These scan data were also used to 
correct different sensitivities among bolometers.  
The pointing observation
was made every 0.5 to 1 hour depending on the weather condition.  The
variation of the pointing is usually within $3''$.  
The data taken under strong wind conditions are discarded because 
of the unstable pointing and the degraded antenna efficiency.
We calibrated the primary flux using the Mars in 1999, and K3-50A and
OH5.89-0.39 in 2000. The flux of the Mars was obtained with the FLUXES
procedure in the STARLINK package, and we employed $6.5\pm 0.2~{\rm
Jy}$ (Sandell 1994) and $8.8\pm 0.9~{\rm Jy}$ (Falcke et al. 1998) for the
fluxes of K3-50A and OH5.89-0.39, respectively.  The beam pattern was
measured using 3C279 which yields $\sigma_{\rm FWHM}=12.\hspace{-2.5pt}''5$ 
and $13.\hspace{-2.5pt}''2$ in 1999 and 2000, respectively. 
An uncertainty of the beam-size is as large as $1''$.  
The beam is slightly elongated along the elevation direction by
$\sim 10\%$, and the side-lobe level amounts to 3\% of the peak value
whereas the Gaussian yields 1\% of the peak value.  
The pointing and gain stabilities
were checked using 1334-127, and the rms variations in the peak fluxes
were 12\% in 1999 and 14\% in 2000.  These variations are largely
ascribed to strong elevation-dependence of the antenna efficiency at
150~GHz.  Correcting this gain variation, we estimate the total
absolute calibration error to be 11\% (7\% in 1999 and 14\% in 2000).  
The larger error in 2000 is due to the fact that the flux of 1334-127 daily
changed up to $\sim 20\%$ probably because of the burst, while it was 
stable in 1999 run. 

Spike noises above the $4\sigma$ level appearing in time-ordered data
were removed. Again the map was created using the PLAIT method; the
$1\sigma$ noise-level in the final image is 1.6~mJy ${\rm beam}^{-1}$
($2.4~{\rm mJy~beam^{-1}}$ in 1999, and $2.1~{\rm mJy~beam^{-1}}$ in
2000).  In the final map, an averaged flux at the edge of map
($64''$ away from the center), $0.50\pm 0.15~{\rm mJy~beam^{-1}}$, was
subtracted to define the zero-level of the data.

\section{Results}

\subsection{Low Angular Resolution Image at 21~GHz}

Figure~1(a) displays the final map of RX~J1347--1145 at 21~GHz, which
shows a bright point source near the center of the cluster.  
To estimate the accurate position and the flux at 21~GHz, we observed
the source with VLA at 8.46~GHz (18~ksec) on May 16, 1999, and at
22.46~GHz (3.6~ksec) on May 20, 1999.
Measured fluxes are $22.42\pm 0.04~{\rm mJy}$ at 8.46~GHz and 
$11.55\pm 0.17~{\rm mJy}$ at 22.46~GHz.
Since the VLA configuration is insensitive to the SZ effect, these
values accurately measure the central radio source flux.
The derived source position is ($13^{\rm h} 47^{\rm m} 30^{\rm
s}\hspace{-5pt}.\hspace{2pt}622 \pm 0^{\rm
s}\hspace{-5pt}.\hspace{2pt}0005$, $-11^\circ 45'
09.\hspace{-2.5pt}''44
\pm 0.\hspace{-2.5pt}''009$), and precisely coincides with that of the
optical center of the central cD galaxy.  Although the X-ray peak
position from the ROSAT/HRI data is offset from the optical center
(Schindler et al. 1997), this offset is smaller than the nominal
pointing uncertainty of ROSAT/HRI. 
In the following discussion, therefore, the X-ray
peak position is assumed to coincide with the optical center and the
central radio source position.  

In total, we have three datasets 
at 21~GHz taken in three years: March 1998 (Komatsu et al. 1999), March 1999
(unpublished), and February and April, 2000 (this paper). The 1998 and
1999 runs were done in the cross-scan mode.  We have measured central
peak-intensities of $8.6\pm 2.0~{\rm mJy~beam^{-1}}$ in 1998, $7.8\pm
1.2~{\rm mJy~beam^{-1}}$ in 1999, and $8.3\pm 0.9~{\rm mJy~beam^{-1}}$
in 2000.  
These results show no significant time variation of the source flux value
during the past two years, and the averaged peak-intensity is $8.2\pm
0.7~{\rm mJy~beam^{-1}}$.

Figure~1(b) plots the 21~GHz map after subtracting the contribution of
the point source, adopting the VLA flux at 22.46~GHz.  Its overall
shape around the central part is fairly similar to that of the X-ray surface
brightness contours overlaid in white solid lines, supporting the SZ
interpretation over the entire cluster scale.  The resulting central
intensity (smoothed over the beam-size) of the SZ decrement relative
to the edge of map ($200''$ away from the center) is $I_{\rm
SZ}(0)=-3.3\pm 0.9~{\rm mJy~beam^{-1}}$, or equivalently, $\Delta
T_{\rm RJ}(0)=-1.6\pm 0.4~{\rm mK}$ in terms of the Rayleigh--Jeans
brightness temperature decrement.  Deconvolving the beam-pattern by
approximating the cluster profile with the spherical isothermal
$\beta$-model with the ROSAT/HRI best-fit values for $\theta_{\rm c}$ and
$\beta$ (\S 1), we obtain the central $y$-parameter, $y(0)=(7.8\pm
2.1)\times 10^{-4}$ (relative to the edge of map). Moreover, if we
average the measured peak-intensities over the datasets taken in 
three years, we obtain $y(0)=(7.7\pm 1.6)\times 10^{-4}$.

Our $y(0)$ agrees very well with the value
expected from the X-ray best-fit parameters, 
$y(0)h_{50}^{1/2}=8.0\times 10^{-4}$, if the SZ effect is negligible 
at the edge of map. 
Note that Komatsu et al. (1999) possibly overestimated the SZ intensity 
at the cluster center by $1.75~{\rm mJy~beam^{-1}}$, as they used the 
central source flux of 13.3~mJy at 21~GHz on the basis of the
power-law interpolation of the spectrum at 1.4, 28.5, and 100~GHz.
We now realize that it is not a good approximation.

\subsection{High Angular Resolution Image at 150~GHz}

Figure~1(c) shows the image at 150~GHz smoothed with a Gaussian
filter so that the effective beam-size becomes $20.\hspace{-2.5pt}''6$ 
(the $1\sigma$ noise-level is $1.3~{\rm mJy~ beam^{-1}}$).  Although the
central point source is fainter at 150~GHz than at 21~GHz, it still
affects the diffuse negative intensity field around the center to some
extent.  The most remarkable feature in this image is a strong
negative intensity region located $\sim 20''$ south-east from 
the center. The peak intensity in this region is 
$-5.4~{\rm mJy~beam^{-1}}$ relative to the edge of map. 
This value is $4.2\sigma$ significance level, and 2.5 times larger 
than that expected from the spherical isothermal $\beta$-model, 
$I_{\rm SZ}(\Delta\theta=20'')\sim -2~{\rm mJy~beam^{-1}}$.

In the following, we quantify the statistical significance of 
the detection of this negative excess emission.
As shown in figure~2, we divide the data into four regions: 
south-east (SE), south-west (SW), north-east (NE), and north-west (NW).
The central region is excluded 
so as to remove the contamination of the point source. 
Then we evaluate fluxes in those regions. The results are summarized 
in table~1 together with the $1\sigma$ error of 2.0~mJy each.
The mean flux averaged over all regions is $-6.4\pm 1.0$~mJy, and thus 
the detection of the SZ decrement at 150~GHz is $6.4\sigma$ level. 
The reality of the negative excess in the SE region is supported
by the following facts; (i) the fluxes of the other three regions are
consistent with each other within the $1\sigma$ level, 
(ii) relative to the mean flux over the other three regions, 
$\bar{F}=-4.7\pm1.15~{\rm mJy}$, the excess flux in the SE region is 
$F_{\rm SE}-\bar{F}=-6.6~\pm 2.3~{\rm mJy}$, 
corresponding to $2.9\sigma$ significance, 
and (iii) the excess in the SE region is persistent both in
1999 and in 2000. Thus we interpret this negative excess flux 
being due to the enhanced SZ effect.

\setcounter{figure}{1}
\begin{figure}
  \begin{center}
    \hspace{-1cm}
    \leavevmode\epsfxsize=6.8cm \epsfbox{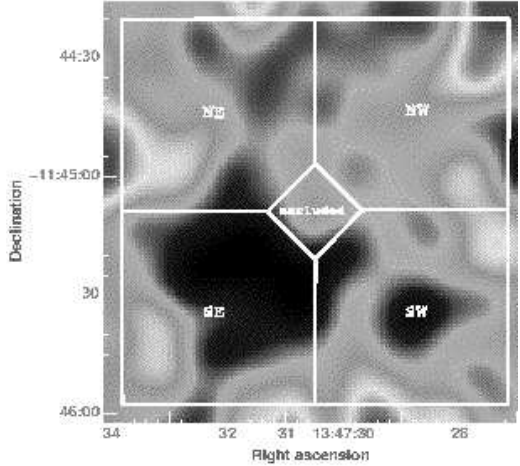}
  \end{center}
\vspace*{-0.5cm}
\caption{
  The definition of regions (see section~3.2).
\label{fig:decomp}
}
\end{figure}

To further explore the significance of this excess, let us consider the
residual map after subtracting the SZ signal assuming the spherical
isothermal $\beta$-model. For this purpose, we employ the best fit
radial profile from the ROSAT/HRI data, and the emission-weighted
temperature, 9.3~keV, measured with ASCA (\S 1).  We refer to
this as model L (low temperature).  After convolving the model image with the
effective beam and correcting for the zero-level, we subtract this 
from the observed map. 
Then we find that the model L implies a central 
radio source flux of $3.8\pm 1.3~{\rm mJy}$, and the expected flux in each
region is $-5.1~{\rm mJy}$. The latter value is consistent with
those listed in table~1 except for the SE region where the difference
is $3.1\sigma$. This is also consistent with the mean flux, 
indicating that the global feature of the SZ effect
at 150~GHz is consistent with model L as well as at 21~GHz.
Figure~1(d) shows the 150~GHz map after subtracting 
the central source flux of 3.8~mJy. While the X-ray contours overlaid 
in the figure seem to trace the detected SZ enhancement to some extent, 
the X-ray flux in the SE region is much smaller than that expected 
from the excess SZ flux we detected at 150~GHz.

In summary, we conclude that we detect an inhomogeneous morphology 
of the SZ signal toward the cluster at 150~GHz. The angular scale of 
the negative excess is around $40''$, and thus cannot be resolved in the
lower angular resolution image at 21~GHz.

\begin{table}
\small
  Table~1.\hspace{4pt}
  Fluxes in four regions toward RX~J1347--1145
  (see figure~2 for the definition).
  Models L and H represent the expected fluxes from 
  the spherical isothermal $\beta$ model with 
  temperatures of $k_{\rm B}T_{\rm e}=9.3~{\rm keV}$ and 
  16.2~keV, respectively.
\begin{center}
  \vspace{6pt}
  \begin{tabular}{p{3.5cm}l}
  \hline\hline
   & flux [mJy] \\
  \hline 
  SE (south-east) \dotfill & $-11.3\pm 2.0$ \\
  NE (north-east) \dotfill & $-4.7\pm 2.0$ \\
  NW (north-west) \dotfill & $-3.3\pm 2.0$ \\
  SW (south-west) \dotfill & $-6.1\pm 2.0$ \\
  mean            \dotfill & $-6.4\pm 1.0$ \\
  \hline
  model L (9.3~keV)  \dotfill & $-5.1$ \\
  model H (16.2~keV) \dotfill & $-8.8$ \\
  \hline
  \end{tabular}
\end{center}
\end{table}

\section{Comparison with Previous Work}

Using the IRAM/Diabolo bolometer array, Pointecouteau et al. (1999)
found that the peak position of the SZ decrement at 143~GHz is offset
to the east of the X-ray peak position.  Since they scanned in a
narrow strip ($30''$ in declination and $120''$ in right ascension)
along west-east direction whose width is comparable
to the beam-size ($\sigma_{\rm FWHM} =23''$), the resulting map is
insensitive to the offset in north-south direction. Thus their
result does not contradict our finding on the image basis.
They presented a different interpretation for this negative
enhancement; they assume that the offset between the X-ray peak 
position and the radio source position is real, and ascribe the 
offset between the SZ and the X-ray peak positions to the contamination 
of the positive radio source embedding in the SZ decrement tracing 
the X-ray signal.  More specifically, they found the best-fit values 
for the central $y$-parameter, $y(0)=12.7^{+2.9}_{-3.1}\times 10^{-4}$, 
and for the central point source flux, $6.1^{+4.3}_{-4.8}~{\rm mJy}$, 
at 143~GHz, adopting the $\beta$-model radial profile from the X-ray data.
This $y(0)$ corresponds to the much higher temperature of 16.2~keV 
than the ASCA value. We call this set of parameters model H 
(high temperature).
When subtracting the SZ flux of the model H from our map at 150~GHz, 
we find the central source flux of $6.6\pm 1.3~{\rm mJy}$. 
Since this value is close to their fit, our data at 150~GHz are 
consistent with their data, apart from the interpretation.

Here we compare our data to theirs quantitatively.
(i) Taking into account the relativistic correction to the SZ effect
(Itoh, Kohyama, Nozawa 1998) to extrapolate their $y(0)$ at 143~GHz 
into the value at 21~GHz, model H predicts 
$y(0)=13.9^{+3.2}_{-3.4}\times 10^{-4}$ at 21~GHz. This value 
significantly exceeds our observed value, 
$(7.7\pm1.6)\times 10^{-4}$, which is well consistent with model L. 
(ii) Model H predicts the flux of $-8.8$~mJy for each
region defined in figure~2, which is systematically smaller than
our observed values listed in table~1. 
Therefore, we conclude that model L with the excess SZ effect is 
more consistent with our data at 21~GHz and 150~GHz data than model H.
Their higher value of $y(0)$ than ours is probably because of their
narrower field-of-view. The excess SZ effect detected in 
the SE region would dominate the mean signal in their map, 
resulting in an overestimate of $y(0)$.

Incidentally the negative flux in the SE region at 150~GHz should
show up as a {\it positive} SZ flux of 11~mJy in the JCMT/SCUBA band
(350~GHz).  While Komatsu et al. (1999) did not identify the
corresponding peak in their SCUBA map, this is not inconsistent one 
another because of the high noise-level in their SCUBA map
($1\sigma=8~{\rm mJy~beam^{-1}}$).

\section{Discussion}

Possible physical explanations for the origin of the excess SZ feature
include a) a projection contamination of another higher redshift
cluster, b) warm gas associated with large-scale filamental
structures, c) a substructure in the cluster gas, d) a cooling flow
around the central region, and e) non-gravitational heating from AGNs and/or
supernovae. Actually any explanation needs to be somewhat contrived,
as it should be simultaneously consistent with the fairly smooth
X-ray brightness distribution. Specifically, a) requires that the
background cluster should be at $z>3$, b) is viable only if the low
temperature ($\sim 0.4$~keV) gas extends over 1~Gpc along the line of
sight, c) implies that the temperature of the substructure is larger
than 200~keV, and d) and e) indicate that either pressure or virial
equilibrium of the intracluster gas should be abandoned.  Thus none of
those possibilities seems to be sufficiently satisfactory.

Nevertheless if the complex morphological structure is generic to other
high redshift clusters, distance measurements to the clusters based on the 
SZ data should be interpreted with caution.  
To elucidate this, let us consider how the enhanced decrement region 
at 150~GHz systematically affects the estimate of the Hubble constant 
from the 21~GHz data.  
The excess SZ flux at 21~GHz in the SE region is expected as $-0.43$~mJy,
based on the deviation from model L at 150~GHz in the same region
($-6.2$~mJy).
Since the extent of the SE region is smaller than the
beam-area at 21~GHz, this flux amounts to 13\% of $I_{\rm SZ}(0)$ at
21~GHz.  This corresponds to overestimating $I_{\rm SZ}(0)$ relative
to the isothermal $\beta$-model prediction, and thus to a systematic
underestimate of $H_0$ by 22\% through the relation 
$H_0\propto I_{\rm SZ}^{-2}(0)$.  
This consideration might be relevant in understanding the
Hubble diagram from the SZ effect (e.g., Kobayashi, Sasaki, Suto 1996;
Birkinshaw 1999).

As demonstrated here, single-dish measurements of the SZ effect with
high angular resolution play a complementary role to
interferometers in exploring the intracluster gas state.
In addition, a more accurate and higher resolution imaging observation
 in X-ray band with {\it Chandra} and {\it XMM--Newton} observatories 
is important to understand the physical processes in this cluster 
as well as the future follow-up SZ observations including
the SZ dedicated interferometers and JCMT/SCUBA.

\par
\vspace{1pc}\par

We thank Akihiro Sakamoto and the NRO staff for their help during the
observation at NRO, Izumi Ohta for observing RX~J1347--1145 with NMA, 
and Wolfgang Reich for providing the data of
OH5.89-0.39 and for useful comments on the data analysis. 
We also thank John Carlstrom, Uro$\check{\rm s}$ Seljak, and 
David N. Spergel for many valuable comments which have improved 
this {\it Letter}.  E.K., T.K. and K.Y. acknowledge 
fellowships from Japan Society for the Promotion of Science.  
This research was supported in part by the Grants-in-Aid for the 
Center-of-Excellence (COE) Research of the Ministry of Education, 
Science, Sports, and Culture of Japan to RESCEU (No. 07CE2002),
and Grand-in-Aid of the Ministry of Education, Sports, and Culture
of Japan (No. 11440060).

\section*{References}
\small
\re
Allen S.W., Fabian A.C.\ 1998, MNRAS 297, L57
\re
Bahcall N.A.\ 1988,
ARA\&A 26, 631
\re
Birkinshaw M.\ 1999, 
Phys. Rep. 310, 97
\re
Carlstrom J.E., Joy M.K., Grego L.\ 1996,
ApJL 456, L75
\re
Carlstrom J.E., Joy M.K., Grego L., Holder G.P., Holzapfel W.L.,
Mohr J.J., Patel S., Reese E.D.\ 2000, 
Physica Scripta T85, 148 
\re
Emerson D.T., Gr\"ave R.\ 1988,
A\&A 190, 353
\re
Falcke H., Goss W.M., Matsuo H., Teuben P., Zhao J.H., Zylka R.\ 1998,
ApJ 499, 731
\re
Inagaki Y., Suginohara T., Suto Y.\ 1995, PASJ 47, 411
\re
Itoh N., Kohyama Y., Nozawa S.\ 1998, ApJ 502, 7
\re
Jing Y.P., Suto Y.\ 2000, ApJL 529, L69
\re
Kobayashi S., Sasaki S., Suto Y.\ 1996, PASJ 48, L107
\re
Komatsu E., Kitayama T., Suto Y., Hattori M.,
Kawabe R., Matsuo H., Schindler S., Yoshikawa K.\ 1999,
ApJL 516, L1
\re
Kuno N., Matsuo H., Mizumoto Y., Lange A.E., Beeman J.W., Haller E.E.\
1993,
Int. J. Infrared Millimeter Waves, 14, 749
\re
Ott M., Witzel A., Quirrenbach A., Krichbaum T.P., Standke K.J.,
Schalinski C.J., Hummel C.A.\ 1994,
A\&A 284, 331
\re
Pointecouteau E., Giard M., Benoit A., D\'{e}sert F.X.,
Aghanim N., Coron N., Lamarre J.M., Delabrouille J.\ 1999,
ApJL 519, L115
\re
Rephaeli .\ 1995,
ARA\&A 33, 541
\re
Sandell G.\ 1994,
MNRAS 271, 75
\re
Schindler S., Hattori M., Neumann D.M., B\"ohringer H.\ 1997,
A\&A 317, 646
\re
Schindler S.\ 1999, 
A\&A 349, 435
\re
Yoshikawa K., Itoh M., Suto Y.\ 1998,
PASJ 50, 203
\re
Zel'dovich Ya.B., Sunyaev R.A.\ 1969,
Ap\&SS 4, 301

\label{last}
\onecolumn

\setcounter{figure}{0}
\begin{figure}[ht]
  \begin{center}
    \leavevmode\epsfxsize=18cm \epsfbox{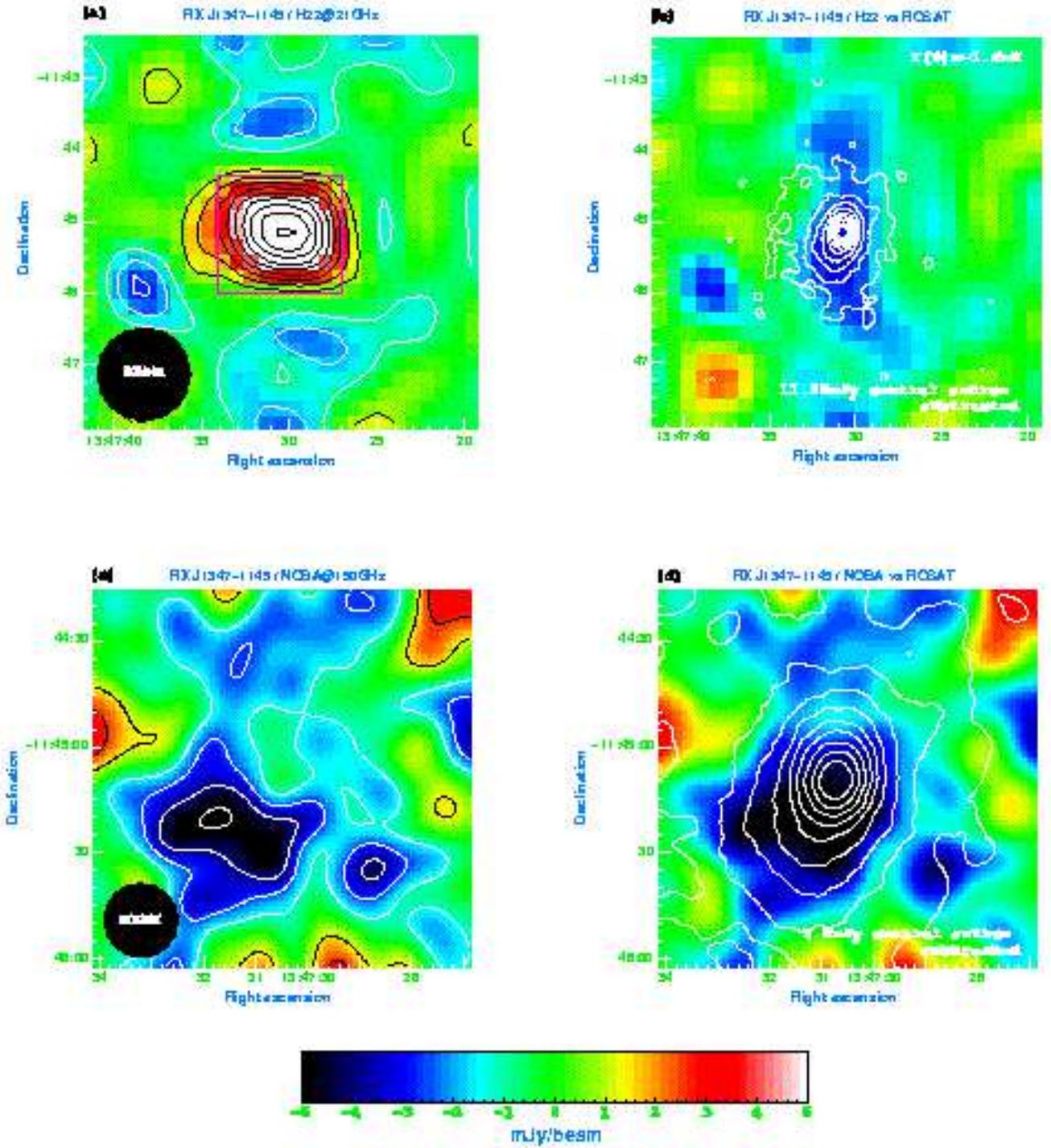}
  \end{center}
\caption{
  The SZ maps of RX~J1347--1145.
  (a) the 21~GHz map before subtracting the central point source.
  Field-of-view is $6'\times 6'$ 
  ($2.4~h_{50}^{-1}~{\rm Mpc}\times 2.4~h_{50}^{-1}~{\rm Mpc}$).
  Black (white) lines represent $1\sigma$ contours of positive  (negative) 
  intensities, where $1\sigma = 0.9~{\rm mJy~beam^{-1}}$.
  The beam-size is shown as a filled black circle of a diameter
  $76.\hspace{-2.5pt}''5$. The area defined by a magenta solid line 
  corresponds to (c) and (d).  
  (b) the 21~GHz map after subtracting the central point source. 
  We used the VLA flux of 11.55~mJy. Overlaid in white lines are 
  the X-ray contours, and represent 10--90\% of the peak value in 10\% 
  intervals.  The equivalent Rayleigh--Jeans temperature at
  the center is $\Delta T_{\rm RJ}(0)=-1.6\pm 0.4~{\rm mK}$.
  (c) the 150~GHz map before subtracting the
  central point source.   Field-of-view is $1'9\times 1'9$ 
  ($0.75~h_{50}^{-1}~{\rm Mpc}\times 0.75~h_{50}^{-1}~{\rm Mpc}$).
   The lines have the same meaning as in the panel
  (a).  In this panel, the beam-size is $20.\hspace{-2.5pt}''6$, and
  $1\sigma = 1.3~{\rm mJy~beam^{-1}}$.
  (d) the 150~GHz map after subtracting the central point source
  (assuming the flux of 3.8~mJy) overlaid with the X-ray contours.
\label{fig:szmap}
}
\end{figure}
\end{document}